\documentclass[12pt]{iopart}

\usepackage{cite}
\usepackage{xcolor}
\usepackage{bbold}
\usepackage{iopams}  
\begin{document}

\title[Interrelation of nonclassicality conditions through stabiliser group homomorphism]{Interrelation of nonclassicality conditions through stabiliser group homomorphism}
  \author{Sooryansh Asthana}
\address{Department of Physics, Indian Institute of Technology Delhi, New Delhi-110016, India.}
\ead{sooryansh.asthana@physics.iitd.ac.in}
\vspace{10pt}
\begin{indented}
\item[]January 2022
\end{indented}

\begin{abstract}
In this paper, we show that coherence witness for a single qubit itself yields conditions for  nonlocality and entanglement inequalities for multiqubit systems. It also yields a condition for quantum discord in two--qubit systems. It is shown by employing homomorphism among the stabiliser group of a single qubit and those of multi--qubit states. Interestingly,  globally commuting  homomorphic images of single qubit stabilisers do not allow for consistent assignments of outcomes of local observables.  As an application,  we show that CHSH inequality can be straightforwardly generalised to nonlocality inequalities for multiqubit GHZ states. It also reconfirms the fact that quantumness prevails even in the large $N$--limit, if coherence is sustained. The mapping provides a way to construct many nonlocality inequalities, given a {\it seed} inequality. This study gives us a motivation to gain a better control over multiple degrees of freedom and multi-party systems. It is because in multi-party systems, the same nonclassical feature, {\it viz.}, coherence may appear in many avatars. 
\end{abstract}

\vspace{2pc}
%
\noindent\submitto{\NJP}
%
\maketitle
%
%

\section{Introduction}

Quantum mechanics has a number of distinctive features, e.g., quantum coherence \cite{Streltsov17}, nonlocality \cite{Brunner14}, steering \cite{Wiseman07steering}, quantum entanglement \cite{Horodecki09}, and, quantum discord \cite{Ollivier01}, etc. They act as resources in quantum--computing \cite{Deutsch92, Shor94, Bernstein97}, quantum--search \cite{Grover96, Grover97}, and, quantum communication algorithms \cite{Bennett92, Bennett93, Bennett01}. Owing to their resource--theoretic importance, recent times have witnessed an unprecedented surge of interest in the  detection and quantification of these features. Despite being closely related, there are subtle differences in the definitions of different nonclassical features. For example, the definition of entanglement (separability) a--prioi assumes quantum mechanics, whereas the assumption of locality is independent of quantum mechanics. Quantum coherence underlies all the nonclassical correlations, be it entanglement or nonlocality or quantum discord. In fact, the interrelations between coherence, entanglement and quantum discord have already been studied from different viewpoints \cite{streltsov2015measuring, adhikary2020bell,  asthana2021non}. 

Nonlocality, in particular, plays a pivotal role in various quantum communication tasks, such as device--independent quantum key distribution \cite{Vazirani14} and randomness generation \cite{Gomez18}, etc. Thanks to its crucial applications,  several approaches have been employed to derive nonlocality inequalities for numerous families of states such as graph states, cluster states, etc. (see, for example, \cite{Brunner14, Baccari20, Niu21} and references therein). Many nonlocality inequalities have been derived by employing stabiliser formalism, logical qubits \cite{Hsu21} and by employing the action of abelian and non--abelian groups \cite{Guney14, Guney15}. Additionally, a procedure for lifting Bell inequalities have also been proposed \cite{pironio2005lifting}. A natural question in this program is how new nonlocality inequalities can be obtained and how the new ones are related to the already existing ones. Stabiliser formalism has also been studied in other contexts, most extensively in quantum error--correcting codes \cite{Gottesman97},  and, in finding entanglement witnesses \cite{Toth05}. 

In this work, we study the interrelation of nonclassicality features in Hilbert spaces of different dimensions. We first show that there exists a homomorphic mapping between stabiliser groups of a single qubit state and an $N$--qubit state.  A key feature that we employ is the existence of several stabiliser operators  of an $N$--qubit system, homomorphic to a single stabiliser of an $M$--qubit system ($M <N$).  We harness it to establish a relation between conditions for coherence in a single qubit, and, those for nonclassical correlations (e.g., nonlocality, entanglement, and quantum discord \cite{Ollivier01}) in multiqubit systems. Different stabilisers of an $N$--qubit system are  products of locally noncommuting operators. This gives us a hint as to why these operators might play a crucial role in  systematic construction of nonlocality inequalities. Locally noncommuting observables are required to detect entanglement as well as nonlocality. That is why we choose  sets of those lower dimensional stabilisers, whose image(s) involves locally noncommuting observables \cite{bavaresco2018measurements}. For the emergence of a nonlocality inequality, a natural requirement is that the inequality gets satisfied purely by classical probability rules, without taking recourse to quantum mechanics \cite{Fine82}. Additionally,  it should also get violated by (at least) one quantum mechanical state.  In particular, while deriving nonlocality inequalities, a key aspect is the identification of observables leading to violation of a constraint put by local--hidden--variable (LHV) models.  It is facilitated via homomorphism between the stabiliser groups.  
This feature provides us with a prescription for constructing a series of nonlocality inequalities, given a seed nonlocality inequality by employing the homomorphic mapping among the stabiliser groups in the reverse direction.

The central results of the paper are as follows. (I) We start by showing how various  nonlocality inequalities involving dichotomic observables emerge from CHSH inequality \cite{Clauser69}. (II)  Entanglement inequalities for the simplest systems, i.e., two--qubit systems, yield several nonlocality inequalities for multiqubit systems, thanks to the homomorphism between stabiliser groups of a two--qubit system and that of a multi--qubit system. (III) This approach naturally   unravels interconnection among various two-qubit  entanglement inequalities and multi--qubit nonlocality inequalities. (IV)  Coherence leads to various nonclassical correlations depending on which homomorphic image of the stabiliser of a single qubit is employed.  It turns out that a less stringent coherence witness and a more stringent  coherence witness are required to obtain conditions for entanglement and nonlocality respectively.

 The paper is organised as follows: in section (\ref{Notation}), we set up the notation to be used in the paper. In section (\ref{Framework}), we set up the framework. Section (\ref{Application}) is central to the paper, in which the emergence of nonlocality inequalities has been shown. In section (\ref{Previous_work}), we discuss the relation of this work with the previous works. Section (\ref{Conclusion}) concludes the paper with closing remarks.

 \section{Notation}
 \label{Notation}
In this section, we set up  compact notations to be used henceforth in the paper for expressing the results. 
\begin{enumerate}
\item   For an  $N$-- party system, the observables $A_i, B_i$ refer to the $i^{\rm th}$ subsystem. Observables belonging to the same subsystem are distinguished by primes in the superscript, such as $A'_i,~A''_i,\cdots$. 
\item The symbols $X_i, Y_i, Z_i$ are reserved for Pauli matrices acting over the space of the $i^{\rm th}$ qubit.
\end{enumerate}
\section{Framework} 
\label{Framework}
In this section, we elaborate on the framework to be used throughout the paper. Suppose that there is a single qubit,
\begin{eqnarray}
    |\psi\rangle = \frac{1}{\sqrt{2}}(|0\rangle +|1\rangle).
\end{eqnarray}
Its stabiliser group is $\{\mathbb{1}_2, X\}$, where $X= |0\rangle\langle 1|+|1\rangle\langle 0|$. Instead, if we assume that both $|0\rangle$ and  $|1\rangle$ are logical qubits, i.e.,
\begin{eqnarray}
    |\psi_L\rangle = \frac{1}{\sqrt{2}}(|0_L\rangle+|1_L\rangle)
\end{eqnarray}
and $|0_L\rangle \equiv |00\rangle$ and $|1_L\rangle \equiv |00\rangle$. The stabiliser group of $|\psi_L\rangle$ will be $\{\mathbb{1}_L, X_L\}$, where $X_L \equiv |0_L\rangle\langle 1_L|+|1_L\rangle\langle 0_L|$. The crux of the matter is that the actions of both the operators-- $\mathbb{1}_L$ and $X_L$-- on the state $|\psi_L\rangle$ are identical to the tensor products of the following local operators,
\begin{eqnarray}
\label{Homo}
    \mathbb{1}_L \rightarrow \mathbb{1}_4, Z_1Z_2;~~X_L \rightarrow X_1X_2, -Y_1Y_2.
\end{eqnarray}
Here $X_i, Y_i, Z_i$ are Pauli operators acting over the space of $i^{\rm th}$ qubit ($i=1, 2$).

This observation can be succinctly formulated in terms of homomorphism among stabiliser groups, which forms the basis of the framework used throughout the paper. The framework essentially hinges on homomorphism between the stabiliser groups of $M$ and $N$--qubit systems ($M <N$).  The homomorphism entails sets of many `locally noncommuting' operators, which have identical actions on an entangled state.  We plan to employ homomorphism of stabiliser groups in the reverse direction for the case of Pauli observables throughout this paper.  We illustrate it with the simplest example of a single qubit and a two--qubit state.

\subsection{Homomorphic mapping between the stabiliser groups}
Consider a single--qubit state,
\begin{eqnarray}
 |\psi\rangle =   \frac{1}{\sqrt{2}}(|0\rangle+|1\rangle), 
\end{eqnarray}
whose stabiliser group is ${G} \equiv \{\mathbb{1}_2, X\}$. Here, the symbol $\mathbb{1}_2$ represents a $2\times 2$ identity matrix and $X$ represents Pauli $X$ operator. This set forms a group under matrix multiplication. 

This group bears a homomorphism with the stabiliser group, $\{\mathbb{1}_4, X_1X_2, -Y_1Y_2, Z_1Z_2\}$, of the two-qubit state, $\frac{1}{\sqrt{2}}(|00\rangle + |11\rangle)$. The homomorphic mapping is given by,
\begin{eqnarray}
    \{X_1X_2, -Y_1Y_2\} &\mapsto X\nonumber\\
    \label{Homo2}
    \{\mathbb{1}_4, Z_1Z_2\} &\mapsto \mathbb{1}_2.
\end{eqnarray}
Moving on to the Hilbert space of three qubits  (i.e., ${\cal H}^2\otimes {\cal H}^2\otimes {\cal H}^2$), the group $G$
bears a homomorphism with the group,
\begin{eqnarray}
  \{ &\mathbb{1}_8, X_1X_2X_3, -X_1Y_2Y_3, -Y_1X_2Y_3, -Y_1Y_2X_3, Z_1Z_2, Z_2Z_3, Z_3Z_1\}.  \nonumber
\end{eqnarray}
   The mapping is given by,
   \begin{eqnarray}
   & \{\mathbb{1}_8, Z_1Z_2, Z_1Z_3, Z_2Z_3\}\mapsto \mathbb{1}_2\nonumber\\
       \label{Homomorphism}
  & \{X_1X_2X_3, -X_1Y_2Y_3, -Y_1X_2Y_3, -Y_1Y_2X_3\} \mapsto X.
   \end{eqnarray}
   Another homomorphic mapping of the group,
   \begin{eqnarray}
  \{ &\mathbb{1}_8, Y_1X_2X_3, -Y_1Y_2Y_3, X_1Y_2X_3, X_1X_2Y_3, Z_1Z_2, Z_2Z_3, Z_3Z_1\},  
\end{eqnarray}
to the group $\{\mathbb{1}_2, Y\}$ is as follows:
   \begin{eqnarray}
   \label{Homomorphism_2}
   & \{\mathbb{1}_8, Z_1Z_2, Z_1Z_3, Z_2Z_3\}\mapsto \mathbb{1}_2\nonumber\\
 & \{Y_1X_2X_3, -Y_1Y_2Y_3, X_1Y_2X_3, X_1X_2Y_3\}\mapsto Y.
   \end{eqnarray}
     Its extension to the  stabiliser group of an $N$-- qubit system is straightforward. An important point is that homomorphic images of single qubit stabilisers are locally noncommuting. So, they can be naturally employed for constructing nonlocality inequalities. In this way, this mapping plays a crucial role in, (i) identifying the forms of observables while constructing a nonlocality inequality, and, (ii) showing the interrelation between coherence witness for a single  qubit, entanglement witness for a two-qubit system, and nonlocality inequalities for $N$-qubit system ($N \geq 3$). The observables are provided by this mapping and the bound can be set by considering the notion of classicality (e.g., LHV models, separability, etc.).
\subsection{Rationale underlying choice of stabiliser group elements yielding nonlocality inequalities}
\label{Rationale_ent}
Since we plan to obtain nonlocality inequalities starting from group homomorphism, it is worthwhile to discuss beforehand the relationale underlying the choices of observables for deriving such inequalities, which is given below:
\begin{enumerate}
\item Nonlocality  in a multi-party system refers to correlations between outcomes of one set of locally non-commuting observables on the first party with another set of locally non-commuting observables on the second party. Guided by this, we choose those elements of the stabiliser group,  which ensure that the ensuing inequality involves the correlations, such as  $\langle A_1A_2\rangle$ and $\langle A'_1A'_2\rangle$, with the proviso that the commutators $[A_1, A'_1] \neq 0; [A_2, A'_2] \neq 0$ (please note that the observables $A_1, A'_1$ and $A_2, A'_2$ act over the spaces of the first and the second party respectively.).

\item A further requirement is, of course, that, the local hidden variable bound should be put such that all the LHVs satisfy the inequality.
\end{enumerate}
We next give the rationale for the choice of observables for constructing entanglement inequalities. 
\subsection{Rationale underlying choice of stabiliser group elements yielding entanglement inequalities}
\label{Rationale_ent_2}
We choose the observables through stabiliser group homomorphism, keeping the following considerations in mind:
\begin{enumerate}
\item  We choose those elements of a stabiliser group,  which ensure that the ensuing inequality involves the correlations, such as  $\langle A_1A_2\rangle$ and $\langle A'_1A'_2\rangle$, with the proviso that the commutators $[A_1, A'_1] \neq 0; [A_2, A'_2] \neq 0$ (please note that the observables $A_1, A'_1$ and $A_2, A'_2$ act over the spaces of the first and the second party respectively.).

\item A further requirement is, of course, that, the  bound on the inequality should be put such that all the separable states satisfy the inequality.
\end{enumerate}


\section{Application: coherence witness $\mapsto$ entanglement witnesses $\mapsto$ nonlocality inequalities}
\label{Application}
This section contains the central results of the paper. In this section, we start by showing how a coherence witness for a single qubit system gets mapped to an entanglement witness for a two--qubit system and to a nonlocality inequality for a three-qubit (more generally, a tripartite) system. 
\subsection{Coherence witness $\rightarrow$ Entanglement witness I}
Consider the  coherence witness,
\begin{eqnarray}
\label{Coherence_witness}
0\leq \langle X\rangle \leq \frac{1}{2}.    
\end{eqnarray}
 We look at the possible homomorphisms to the group of stabilisers of two-qubit Bell state and show that the entanglement inequalities and conditions for quantum discord follow. The coherence witness, $\langle X\rangle \leq \frac{1}{2}$, and a coherent state, $\frac{1}{\sqrt{2}}(|0\rangle+|1\rangle)$, get mapped to  an entanglement witness, $\langle X_1X_2-Y_1Y_2\rangle \leq 1$, and to an entangled state, $\frac{1}{\sqrt{2}}(|00\rangle+|11\rangle)$ respectively, under the homomorphism given in equation (\ref{Homo2}), i.e.,
 \begin{eqnarray}
     &\langle{X}\rangle \leq \frac{1}{2} \Rightarrow 2\langle X\rangle \leq 1\nonumber\\
  & 2\langle X\rangle \leq 1   \mapsto  \langle X_1X_2-Y_1Y_2\rangle \leq 1,\nonumber\\
  \label{Mapping__}
    & \frac{1}{\sqrt{2}}(|0\rangle+|1\rangle)\mapsto \frac{1}{\sqrt{2}}(|00\rangle+|11\rangle).
 \end{eqnarray}
 Note that we have employed the mapping,  $|0\rangle\mapsto |00\rangle$ and $|1\rangle \mapsto |11\rangle$ in the last line of equation (\ref{Mapping__}). 
\subsection{Coherence witness $\rightarrow$ Entanglement inequality II}
Secondly, the coherence witness, $\langle X\rangle < 0$, maps to the entanglement witness $ \langle \mathbb{1}+X_1X_2+Y_1Y_2+Z_1Z_2 \rangle < 0$ as follows:
\begin{eqnarray}
\label{Cohwit}
    &\langle X\rangle < 0\Rightarrow \langle 1-1+X+X\rangle < 0
    \end{eqnarray}
    Employing the mapping, $\{X_1X_2, Y_1Y_2\} \mapsto X,$ $\{\mathbb{1}_4, -Z_1Z_2\}\mapsto \mathbb{1}_2$, equation (\ref{Cohwit}) assumes the following form:
    \begin{eqnarray}
     \langle\mathbb{1}+X_1X_2+Y_1Y_2+Z_1Z_2 \rangle < 0,
\end{eqnarray}
which is the optimal linear entanglement witness for two--qubit systems \cite{Guhne2003experimental}. In this manner, we observe that choices of observables appearing in two--qubit entanglement witnesses are governed by the observables appearing in coherence witnesses for a single--qubit system. We show, in the subsequent sections, that a similar conclusion holds for nonlocality as well.

\subsection{Coherence witness $\mapsto$ Mermin's nonlocality inequality}
We next move on to show that the same coherence witness, $\langle X\rangle \leq \frac{1}{2}$, maps to Mermin's nonlocality inequality, under the mapping,
\begin{eqnarray}
    |000\rangle\mapsto |0_L\rangle,~|111\rangle\mapsto |1_L\rangle.
\end{eqnarray}
We employ the homomorphism given in equation (\ref{Homomorphism}).
So, the coherence witness, $\langle X\rangle \leq \frac{1}{2}$, maps to the following inequality,
\begin{eqnarray}
   \langle  X_1(X_2X_3-Y_2Y_3)-Y_1(X_2Y_3+Y_2X_3)\rangle \leq 2,
\end{eqnarray}
which is Mermin's inequality \cite{Mermin90}. This procedure admits a straightforward generalisation to a higher number of parties.

At this juncture, it is pertinent to enquire into the special choice of $\frac{1}{2}$ on the right hand side of coherence witness in equation (\ref{Coherence_witness}). This value has been chosen because, the ensuing inequality for a two--qubit turns out to be an entanglement inequality and the one for $N$--qubit system turns out to be Mermin's nonlocality inequality. We next show that any value less than $\frac{1}{2}$ may be used to construct a wintess for quantum discord. 

Secondly, it is also worth--noticing that we have taken all possible homomorphic maps of $X $ in the group $\{\mathbb{1}_2, X\}$. It is because, in order to detect nonlocality/entanglement, it is required to have locally noncommuting operators in the expression.  In the next section, we show how  the condition for quantum coherence for a single qubit and quantum discord for a two--qubit system are related to each other.

\subsection{Coherence witness $\mapsto$ Condition for quantum discord}
Consider a coherence witness $\vert\langle X\rangle\vert \leq \epsilon$ for a single qubit, with $0< \epsilon \leq \frac{1}{2}$. Under the homomorphic mapping $\{X_1X_2, -Y_1Y_2\}\mapsto X$, it gets mapped to two conditions, $\vert\langle X_1X_2\rangle \vert\leq \epsilon$ and $\vert\langle -Y_1Y_2\rangle \vert\leq \epsilon $, which are conditions for quantum discord for a two--qubit system.  The proof is by explicit construction. 
        By definition, a state whose discord, ${\cal D}^{1 \mapsto 2}$ vanishes has the structure,
      \begin{eqnarray}
      \label{non-discordant}
      \rho_{12} = \sum_k p_k|\phi_{1k}\rangle\langle \phi_{1k}|\otimes \rho_{2k},
      \end{eqnarray}
      where, $\sum_k p_k|\phi_{1k}\rangle\langle \phi_{1k}\vert$ is the resolution of $\rho_1$
  in its eigenbasis.

 We choose two orthogonal observables $X_1, Y_1$ for the first qubit with the stipulation that  one of them say, $X_1$ shares its eigenbasis with $\rho_1$ (which is the reduced density matrix of $\rho_{12}$).  The eigen-basis of $Y_1$ is unbiased with respect to that of $X_1$. Thus, a partial tomography is warranted. For  this reason, the condition for discord does not qualify as   a witness.

\subsection{Coherence witness $\mapsto$ CHSH inequality} 
 We now show how CHSH inequality emerges as a descendant of a more stringent coherence witness $\langle X\rangle \leq \frac{1}{\sqrt{2}}$. Under the homomorphic map given in equation (\ref{Homo2}), this inequality gets mapped to,
\begin{eqnarray}
    \langle X_1X_2-Y_1Y_2\rangle \leq \sqrt{2},
\end{eqnarray}
which can be reexpressed as,
\begin{eqnarray}
    \Bigg\langle \Bigg(\frac{X_1-Y_1}{\sqrt{2}}+\frac{X_1+Y_1}{\sqrt{2}}\Bigg)X_2+\Bigg(\frac{X_1-Y_1}{\sqrt{2}}-\frac{X_1+Y_1}{\sqrt{2}}\Bigg)Y_2\Bigg\rangle \leq 2.\nonumber
\end{eqnarray}
We may replace the observables acting over a two--qubit system by generic dichotomic observables $A_1, A_2, A'_1, A'_2$, which renders the seminal CHSH inequality \cite{Clauser69},
\begin{eqnarray}
    \langle (A_1+A'_1)A_2+(A_1-A'_1)A'_2\rangle \leq 2.
\end{eqnarray}
We now move on to show how Mermin's inequality emerges from CHSH inequality by employing group homomorphism.

\subsection{CHSH inequality $\mapsto$ Mermin's inequality}
 We are now in a position to find the descendants of CHSH inequality under different homomorphic maps. Consider, once again, the CHSH inequality \cite{Clauser69},
\begin{eqnarray}
\label{CHSH}
    \langle A_1 (A_2+A'_2)+A'_1(A_2-A'_2)\rangle \leq 2.
\end{eqnarray}
The observables acting over the second qubit may be chosen as, $A_2\equiv X_{2L}, A'_2\equiv Y_{2L}$. We have assumed that the second qubit is logical one and itself composed of two qubits. Employing the group homomorphism between stabiliser group of the second qubit and that of two--qubit states, the observable for the single logical qubit maps to the following tensor products of local observables, 
\begin{eqnarray}
    X_{2L}\mapsto X_2X_3; -Y_2Y_3\nonumber\\
    ~Y_{2L}\mapsto X_2Y_3;Y_2X_3.
\end{eqnarray}
Now, thanks to this mapping, the inequality (\ref{CHSH}) will assume the following form:
\begin{eqnarray}
    \langle A_1(A_2A_3+A'_2A'_3)+A'_1(A_2A'_3-A'_2A_3)\rangle \leq 2,
\end{eqnarray}
which is Mermin's inequality \cite{Mermin90}. It can be further extended for finding nonlocality inequalities for an $N$--party systems similarly. 


\subsection{CHSH inequality $\mapsto$ Das--Datta--Agrawal inequality}
In this section, we show how CHSH inequality for a different choice of observables, coupled with homomorphic map, yields Das--Datta--Agrawal inequality \cite{Das17}. The seminal CHSH inequality \cite{Clauser69} can be written as,
\begin{eqnarray}
    \langle (A_1+A'_1)A_2+(A_1-A'_1)A'_2\rangle \leq 2.
\end{eqnarray}
It gets violated by the state $\frac{1}{\sqrt{2}}(|0\rangle|1\rangle_L-|1\rangle|0\rangle_L)$ for the following choice of observables,

$$A_{1}=\frac{X_{1}+Z_{1}}{\sqrt{2}}, A'_{1}= \frac{X_{1}-Z_{1}}{\sqrt{2}}, A_{2L}=X_{2L}, A'_{2L}=Z_{2L}.$$ 
We may write the logical operators as a tensor product of local operators, i.e.,  $X_{2L}\mapsto X_2X_3$ and $Z_{2L}\mapsto Z_2$. In terms of generic observables, it will lead to the following inequality,
\begin{eqnarray}
    \langle (A_1+A'_1)A_2A_3+(A_1-A'_1)A'_2\rangle \leq 2,
\end{eqnarray}
which is the same as Das--Datta--Agrawal inequality for a tripartite system. In this language, various inequalities, which have been derived through diverse considerations may be regarded as descendants of a smaller set of nonlocality inequalities under different homomorphic maps.

Up to this point, we have considered only the homomorphic mappings to the group $\{\mathbb{1}_2, X\}$ or $\{\mathbb{1}_2, Y\}$ or $\{\mathbb{1}_2, Z\}$ separately or at most, to two of them at a time. We now consider the homomorphic maps of all the three single qubit Pauli operators $X, Y$ and $Z$ simultaneously in the sections to follow. 
\subsection{Multi--party nonlocality inequalities from two--qubit entanglement inequalities}
We are now in a position to start straightaway from two--qubit entanglement witnesses and show the emergent hierarchical structure of inequalities. We can simply assume that one of the two qubits is a logical one. The logical qubit itself is composed of many qubits. The crux of the matter is that the operators appearing in the entanglement inequalities corresponding to the logical qubit become logical operators. These operators may, in turn,  be written as direct products of local operators in more ways than one. 

To illustrate it, in this section, we map the entanglement inequality \cite{Werner89},
\begin{eqnarray}
\label{Ent2qubit}
\langle X_1X_{2L}+Y_1Y_{2L}+Z_1Z_{2L}\rangle \leq 1,  
\end{eqnarray}
 to various nonlocality inequalities for three and four party systems. The symbols $X_{2L}, Y_{2L}, Z_{2L}$ represent that the second qubit may be  a logical one. The inequality (\ref{Ent2qubit}) gets maximally violated by the state $\frac{1}{\sqrt{2}}(|01\rangle-|10\rangle)$. If we assume the second qubit to be a logical one, the state will become $\frac{1}{\sqrt{2}}(|01_L\rangle-|10_L\rangle) \Big(\equiv \frac{1}{\sqrt{2}}(|011\rangle-|100\rangle)\Big)$, i.e., it maps to a three-qubit GHZ state.
 
\subsubsection{First inequality: tripartite state}
We assume that the second logical qubit consists of two qubits. We observe the homomorphic map, $X_2X_3, -Y_2Y_3\mapsto X_{2L}$, $X_2Y_3, Y_2X_3\mapsto Y_{2L}$,  and $ Z_2, Z_3\mapsto Z_{2L}$. Employing this homomorphic map, the left hand side of equation (\ref{Ent2qubit}) maps to,
\begin{eqnarray}
\label{NL1}
    & \langle X_1(X_2X_3-Y_2Y_3)+Y_1(X_2Y_3+Y_2X_3)+Z_1(Z_2+Z_3)\rangle\leq 4.
\end{eqnarray}
The bound of $4$ has been set by considering all the local hidden variable models. The inequality (\ref{NL1}) is a nonlocality inequality as it gets satisfied by all possible LHVs.  Note that unlike in the previous section, the bound has been set by considering all the LHV models. We may easily make a reverse substitution in  inequality (14) to obtain a corresponding condition for entanglement in the effective two--qubit system. It gets maximally violated by the three-qubit GHZ state $\frac{1}{\sqrt{2}}(|011\rangle -|100\rangle)$, whose expectation value is $6$ for the operator given in (\ref{NL1}). The inequality (\ref{NL1}) can be generically written as,
\begin{eqnarray}
    & \langle A_1(A_2A_3-A'_2A'_3)+A'_1(A_2A'_3+A'_2A_3)
    +A''_1(A''_2+A''_3)\rangle\leq 4,
\end{eqnarray}
where $A_i, A'_i$ and $A''_i$ are dichotomic observables, with outcomes $\pm 1$.

At this stage, we wish to point out a difference between homomorphic images of $X, Y$ and $Z$. Though the homomorphic images of $X$ and $Y$ do not allow for consistent assignments of joint outcomes of local observables, those of $Z$ do. We believe that this feature has its genesis in the fact that the Pauli operators $X$ and $Y$ detect coherence in a single qubit state, whereas the operator $Z$ does not (in the computational basis). Since coherence is a nonclassical property, it gets mapped to entanglement and nonlocality in multiqubit systems. This is what gets reflected in the homomorphic maps of $X$ and $Y$, not allowing for joint assignments of outcomes of local observables \footnote{All the statements have been made in the computational basis $\{|0\rangle, |1\rangle\}$, which consists of the eigenstates of $Z$.}.

\subsubsection{Second inequality: four--party state}
The second logical qubit in equation (\ref{Ent2qubit}) may be assumed to be composed of three qubits. Under this assumption, in this section, we replace $X_{2L}, Y_{2L}$ and $Z_{2L}$ in equation (\ref{Ent2qubit}) by their homomorphic images acting over the tensor product space of three--qubits, i.e., ${\cal H}^2\otimes {\cal H}^2\otimes {\cal H}^2$. That is,
we make the following replacement, $X_{2L}\mapsto X_2X_3X_4, -Y_2Y_3X_4, -Y_2X_3Y_4, -X_2Y_3Y_4$, and, $Y_{2L} \mapsto X_2X_3Y_4, X_2Y_3X_4, Y_2X_3X_4, -Y_2Y_3Y_4$ and $Z_{2L} \mapsto Z_2, Z_3, Z_4, Z_2Z_3Z_4$. This substitution yields the following inequality,
\begin{eqnarray}
    \big\langle &X_1(X_2X_3X_4-X_2Y_3Y_4-Y_2X_3Y_4-Y_2Y_3X_4)\nonumber\\
    +& Y_1(Y_2X_3X_4+X_2Y_3X_4+X_2X_3Y_4-Y_2Y_3Y_4)\nonumber\\
    +& Z_1(Z_2 +Z_3+ Z_4+Z_2Z_3Z_4)\big\rangle \leq 8.
\end{eqnarray}
The bound on the right--hand side has been so fixed that the inequality gets satisfied by all the LHVs. It gets violated by the four--qubit GHZ state, for which the expectation value is equal to the algebraic bound, i.e., 12.

This procedure admits a straightforward generalisation to $N$--qubits. At this juncture, it is worth--mentioning that if we use these stabilisers, different variants of Mermin's inequality \cite{Mermin90} and Das--Datta--Agrawal inequality \cite{Das17} emerge naturally. 

We now show how a nonlocality inequality, derived originally for a cluster  state, may be looked upon as a descendant of a three--qubit entanglement witness.

\subsection{Rederivation of Bell's inequality for cluster state and its descendants}
We have, so far, mapped single qubit logical states $|0\rangle_L, |1\rangle_L$ to $|0\cdots 0\rangle, |1\cdots 1\rangle$, i.e., to states having concatenated zeros and ones. A natural question is what happens if we replace single qubit logical states with arbitrary superpositions of multiqubit states. This is not without interest as we observe that in 5--qubit error correction code, the logical qubits are superpositions of multi--qubit states \cite{Knill01}. We give one such example in this section. A nonlcoality inequality, getting maximally violated by  the following four--qubit state,
\begin{eqnarray}
    |\psi\rangle = \frac{1}{2}\{|00\rangle(|00\rangle+|11\rangle)+|11\rangle(|00\rangle-|11\rangle)\},
\end{eqnarray}
has been derived in \cite{Walther05}. We show how it can be looked upon as a descendant of a three--qubit entanglement witness. Assume that, $|0\rangle_L\mapsto \frac{1}{\sqrt{2}}(|00\rangle+|11\rangle)$, $|1\rangle_L\mapsto\frac{1}{\sqrt{2}}(|00\rangle-|11\rangle)$. We now attempt to find the operators having the same effect as $Y_3X_4, X_3Y_4, X_3X_4, Y_3Y_4$ on single qubit states $|0\rangle_L$ and $|1\rangle_L$. The requisite mapping is as follows:
\begin{eqnarray}
 Y_3X_4, X_3Y_4 \mapsto Y_L; X_3X_4, -Y_3Y_4\mapsto -Z_L.   
\end{eqnarray}
 Under this mapping, the entanglement inequality,
\begin{eqnarray}
    \langle X_1X_2X_{3L} +\mathbb{1}_2Z_2Z_{3L}\rangle \leq 1,
\end{eqnarray}
for a three--qubit system gives rise to the following nonlocality inequality,
\begin{eqnarray}
    \langle X_1X_2(X_3Y_4+Y_3X_4)+\mathbb{1}_2Z_2(X_3X_4-Y_3Y_4)\rangle \leq 2.
\end{eqnarray}
Note that in this section, we have assumed that single logical qubits, $|0\rangle_L, |1\rangle_L$ to map to a coherent superposition of two states, {\it viz.}, $|00\rangle$ and $|11\rangle$. \\

\subsection{Multiparty nonlocality inequalities from nonlinear entanglement inequalities}
We now turn our attention to derivation of multiparty nonlinear nonlocality inequalities starting from a two--qubit nonlinear entanglement inequality. A nonlinear entanglement inequality is given as \cite{Guhne06}, 
\begin{eqnarray}
  &  \langle X_{1}X_{2}+Y_{1}Y_{2}+Z_{1}Z_{2}\rangle
    -\frac{1}{2}\Big(\langle{X_{1}+X_{2}}\rangle^2+\langle{Y_{1}+Y_{2}}\rangle^2\Big)\leq 1.
\end{eqnarray}
 In this case, we assume that both the first and the second qubits are logical and composed of three qubits each. We employ the homomorphic mappings of single qubit operators $X, Y, Z$ to three-qubit operators to arrive at the following nonlocality inequality,
\begin{eqnarray}
 &   \Big\langle (X_1X_2X_3-X_1Y_2Y_3-Y_1X_2Y_3-Y_1Y_2X_3)\nonumber\\
 &(X_4X_5X_6-X_4Y_5Y_6-Y_4X_5Y_6-Y_4Y_5X_6)\Big\rangle\nonumber\\
 +&\Big\langle (Y_1X_2X_3+X_1Y_2X_3+X_1X_2Y_3-Y_1Y_2Y_3)\nonumber\\
 &(Y_4X_5X_6+X_4Y_5X_6+X_4X_5Y_6-Y_4X_5X_6)\Big\rangle\nonumber\\
 +&\Big\langle(Z_1+Z_2+Z_3+Z_1Z_2Z_3)
 (Z_4+Z_5+Z_6+Z_4Z_5Z_6)\Big\rangle\nonumber\\
 -\frac{1}{2}&\big[\langle X_1X_2X_3-X_1Y_2Y_3-Y_1X_2Y_3-Y_1Y_2X_3\rangle\nonumber\\
 +&\langle{X_4X_5X_6-X_4Y_5Y_6-Y_4X_5Y_6-Y_4Y_5X_6}\rangle\big]^2 \leq 32
\end{eqnarray}
The LHV bound is  32, whereas quantum mechanical maximum value is 48, which is achieved by six--qubit GHZ state.

\subsection{Descendants of Mermin's inequality}
We know that the tri--partite Mermin's inequality is given as \cite{Mermin90},
\begin{eqnarray}
    \langle (A_1A_2+A'_1A'_2)A_3+(A_1A'_2-A'_1A_2)A'_3\rangle \leq 2.
\end{eqnarray}
The above inequality gets maximally violated by the three-qubit GHZ state, $\frac{1}{\sqrt{2}}(|000\rangle+|111\rangle)$, for the following choices of observables,
\begin{eqnarray}
    A_1\equiv X_1;~ A'_1\equiv -Y_1;~ A_2\equiv X_2;~ A'_2\equiv Y_2;~ A_3\equiv X_3;~ A'_3\equiv -Y_3.
\end{eqnarray}
If we  assume that the third party is a logical qubit and composed of $N$-- qubits, the corresponding operators will have the following forms,
\begin{eqnarray}
    A_{3L} \equiv X_{3L}; A'_{3L}\equiv -Y_{3L}.
\end{eqnarray}
The forms of the logical operators are given below:
\begin{eqnarray}
    X_{3L} &\equiv |0_L\rangle\langle 1_L|+|1_L\rangle\langle 0_L|;~~
      Y_{3L} \equiv -i\big(|0_L\rangle\langle 1_L|-|1_L\rangle\langle 0_L|\big)
\end{eqnarray}
We next find out the products of local operators corresponding to $X_{3L}$ and $Y_{3L}$. We consider the two cases of $|0_L\rangle$ and  $|1_L\rangle$ conprising of  two and three qubits separately.
\subsubsection{Logical qubits comprising of two qubits}
In this case, we employ the following mapping,
\begin{eqnarray}
     \{X_3X_4, -Y_3Y_4\} &\mapsto X_{3L}; \{X_3Y_4, Y_3X_4\} \mapsto Y_{3L},
\end{eqnarray}
to find the descendant inequality of Mermin's inequality. The descendant inequality is given as,
\begin{eqnarray}
    \langle (A_1A_2+A'_1A'_2)(A_3A_4-A'_3A'_4) +(A_1A'_2-A'_1A_2)(A'_3A_4+A_3A'_4)\rangle \leq 4. 
\end{eqnarray}
\subsubsection{Logical qubits comprising of three qubits}
In this case, we employ the following mapping,
\begin{eqnarray}
     \{X_3X_4X_5, -X_3Y_4Y_5, -Y_3X_4Y_5, -Y_3Y_4X_5\} &\mapsto X_{3L},\nonumber\\
     \{-Y_3Y_4Y_5, Y_3X_4X_5, X_3Y_4X_5, X_3X_4Y_5\} & \mapsto Y_{3L}.
\end{eqnarray}
to find the descendant inequality of Mermin's inequality. The descendant of Mermin's inequality is given as,
\begin{eqnarray}
    \big\langle &(A_1A_2+A'_1A'_2)(A_3A_4A_5-A_3A'_4A'_5-A'_3A_4A'_5-A'_3A'_4A_5)\nonumber\\
    +&(A_1A'_2-A'_1A_2)(-A'_3A'_4A'_5+A'_3A_4A_5+A_3A'_4A_5+A_3A_4A'_5)\big\rangle \leq 8.
\end{eqnarray}

In a similar manner, it can be generalised to an arbitrary $N$--qubit system.

\subsection{Descendants of Svetlichny inequality}
In this same way, we can find the descendants of Svetlichny inequality. The tripartite Svetlichny inequality is given as \cite{Svetlichny87},
\begin{eqnarray}
    \langle(A_1A_2+A'_1A'_2)A_3+(A'_1A_2-A_1A'_2)A_3'\rangle\leq 4,
\end{eqnarray}
where $A_1=B_1+B'_1$ and $A'_1 = B_1-B'_1$. It gets maximally violated by the three--qubit GHZ state $\frac{1}{\sqrt{2}}(|000\rangle+|111\rangle)$.  As before, if we assume that the third qubit is a logical qubit and itself consists of three qubits, we may find descendant of Svetlichny inequality for a five party system. The descendant inequality is given by,
\begin{eqnarray}
    \langle (A_1A_2+A'_1A'_2)A_3A_4A_5+(A'_1A_2-A_1A'_2)A'_3A'_4A'_5\rangle\leq 4,
\end{eqnarray}
which is obtained by employing the mapping $A_3\mapsto A_3A_4A_4$ and $A'_3\mapsto A'_3A'_4A'_5$.

This concludes the applications of the proposed approach. In this way, we can employ this approach to find  descendants of any given inequality.

\section{Relation with previous work}
\label{Previous_work}
Stabilisers have been employed to construct nonlocality and entanglement inequalities \cite{Hsu21,Toth05}. Nevertheless, homomorphism among stabiliser groups has not been hitherto studied, to the best of our knowledge, to derive a number of nonlocality inequalities given a seed nonlocality inequality. Furthermore, this approach also shows an interrelation between coherence witnesses for a single qubit and entanglement and nonlocality inequalities for two and multi--qubit systems.

\section{Conclusion}
\label{Conclusion}
  In summary, we have developed a procedure, based on homomorphism of stabiliser groups, that shows interconnection among coherence witnesses of a single qubit system, entanglement witnesses, and nonlocality inequalities of multiparty systems, and condition for quantum discord. We have applied the procedure to a number of  inequalities to find their descendants. In general, this approach can be employed to obtain a series of (multiparty) nonlocality inequalities, contingent on the knowledge of one seed inequality for a bipartite system.  The framework is quite generic. We have considered but some examples.  Its generalisation to qudits forms an interesting study that will be taken up separately. It also shows that more and more substructures of quantumness emerge as we probe more particles that are in coherent superposition. Since multipartite nonlocality may also be regarded as a special case of sequential contextuality in single--party systems \cite{heywood1983nonlocality}, the proposed procedure may also be employed to render a number of contextuality inequalities.
  
  This work also leaves open a lot of questions: (i) whether an arbitrary entanglement witness can be converted into a nonlocality inequality by suitably extending the number of parties?, (ii) whether a very weak (less stringent) coherence witness for a single qubit system be converted to condition for nonclassical correlations in multiparty systems? If yes, how many subsystems are required as a function of bound of coherence witnesses?  The answers to these questions will hopefully improve the understanding of interrelation among nonclassicality criteria and also serve to deliver a number of conditions for different nonclassical correlations. Furthermore, since nonlocality and entanglement inequalities are, by construction, contextuality inequalities, the inequalities derived via this approach are also contextuality inequalities.

  \section*{Acknowledgements}
   I would like to thank  V. Ravishankar
 for motivating to start the work, insightful comments on it and, in particular, for explaining the physical implication of this work with substructures of a composite system, apart from a purely mathematical equivalence. It is a pleasure to thank Rajni Bala for several discussions and insightful comments. Financial assistance from CSIR (Grant no.: 09/086 (1278)/2017-EMR-I) is gratefully acknowledged.

\end{document}